# A scaling law for the cosmological constant from a stochastic model for cosmic structures

**Salvatore Capozziello** [(1)] **and Scott Funkhouser**[(2)]

[(1)]Dipartimento di Scienze Fisiche, Università di Napoli "Federico II" and INFN Sez. di Napoli, Compl. Univ. Monte S. Angelo, Ed. N., Via Cinthia, I-80126 Napoli, Italy
[(2)]Department of Physics, the Citadel, 171 Moultrie St. Charleston, SC 29409, USA

ABSTRACT
A set of scaling laws, based on the stochastic motions of the granular components of astronomical systems, is applied to a cosmological model with a positive cosmological constant. It follows that the mass of the dominant particle in the observable universe must be proportional to the sixth root of the cosmological constant and of order near the nucleon mass, which is consistent with the Zel'dovich scaling law. The approach is a natural way to solve the "cosmic coincidence" problem. On the other hand, the observed value of the cosmological constant emerges as the result of a scaling law induced by the stochastic mechanism which gives rise to the gravitationally bound systems.

## 1. Introduction

The characteristic observed radii of typical self-gravitating objects can be inferred from microscopic fundamental scales through phenomenological scaling laws [1-4]. This problem was posed for the first time by Dirac [5] and then developed by other authors which supposed the origin of coarse-grained astrophysical and cosmological structures starting from quantum mechanical fluctuations [6]. The debate involves the same origin of the gravitational field which either can be the average result of the non-gravitational interactions or, vice versa, the non-gravitational interactions are "induced" from the gravitational field dominating the universe at any scale.

In [1,2], a set of scaling laws have been established in order to relate mutually the basic physical parameters of the predominant astronomical structures and the nucleon. It was shown that the characteristic sizes of astrophysical and cosmological structures, where gravity is the only overall relevant interaction assembling the system, have a phenomenological relation to the microscopic scales whose order of magnitude is essentially ruled by the Compton wavelength of the proton. This result agrees with the absence of screening mechanisms for the gravitational interaction and could be easily connected to the presence of correcting terms to the Newtonian potential which introduces typical interaction lengths. The emergence of such terms is predicted by several alternative theories of gravity aimed to solve the shortcomings of General Relativity (see [13-15] for reviews on the argument).

To start our considerations, let us take into account an astrophysical system, denoted by $j$, that is gravitationally bound, or nearly so, and that consists of some very large number $N_{kj}$ of granular components, denoted by $k$. The principal hypothesis of this *Fluctuative Scaling Law Model* is that the characteristic relaxation-time $t_j$ of the astrophysical system and the characteristic time $t_k$ of the statistical fluctuations associated with any one of the granular components are, on average, related by

$$t_j \sim t_k N_{kj}^{1/2}. \qquad (1)$$

Note that the number $N_{kj}$ is of the order $M_j/M_k$, where $M_j$ and $M_k$ are respectively the masses of the astrophysical system and the granular component. The hypothesis in (1) is based on the simple premise that the dynamics of a system, consisting of a large number of particles, be determined statistically by the stochastic fluctuations of the particles [1-4].

There exists a variety of physical consequences associated with this model, but, for the purposes of the present work, it is necessary to consider only two scaling laws that follow from (1). It follows that the characteristic action $A_j$ of an astrophysical system should be related to the action $A_k$ of the granular component according to

$$A_j \sim A_k N_{kj}^{3/2}. \qquad (2)$$

Furthermore, the characteristic radius $R_j$ of the astrophysical system should be scaled to the characteristic radius $R_k$ of the granular component by

$$R_j \sim R_k N_{kj}^{1/2}. \qquad (3)$$

The ramifications of (1), (2) and (3) are broad. Let the subscripts $n$, $s$, $g$ and $u$ denote respectively the characteristic parameters of nucleon, solar systems, galaxies and the observable universe in this epoch (the term "observable" is used in the literature to refer to the parameters defined in terms of the event horizon and of the particle horizon. In this epoch, both horizons are of the same order, and the term "observable" may apply to both sets of associated parameters in this epoch [11]). Even if the bulk of the masses of galaxies and the observable universe are non-baryonic, according to the hypothesis of dark matter, the difference between $N_{ng}$ and $M_g/M_n$ would be very small with respect to $N_{ng}$, and the difference between $N_{nu}$ and $M_u/M_n$ would be very small with respect to $N_{nu}$. Similarly, the numbers $N_{sg}$, $N_{su}$ and $N_{gu}$ would differ from their respective approximations $M_g/M_s$, $M_u/M_s$ and $M_u/M_g$ by relatively small amounts if dark matter dominates the masses of the observable universe and galaxies. The observable universe, galaxies, solar systems and nucleons constitute therefore a *hierarchy* in which any given body is, approximately, either an aggregate of other bodies within the hierarchy or a granular component constituting another body within the hierarchy. As such, the characteristic parameters of those bodies should be related approximately through the scaling laws in (1), (2) and (3). The mutual relationships for the nucleons, solar systems, galaxies and the observable universe are consistent with observations, which constitutes a strong validation for the physical meaning of such fluctuative scaling law model [1,2].

While the characteristic parameters of galaxies, solar systems and particles are presumably fixed, the event horizon and the particle horizon, which represent the only two measures of the characteristic size of the universe, change significantly during the evolution of the universe [11]. It is therefore remarkable that the current parameters of the observable universe satisfy the relationships specified by the model. Disregard momentarily the knowledge that the current parameters are empirically consistent with the model in [1,2] and consider the evolution of the universe from the Big Bang to some point in the distant future including the structure formation epoch and the present dark energy epoch. If the fluctuative scaling law model is physically meaningful, then there should exist some set of characteristic cosmic parameters that represents the manifestation of the stochastic behaviors associated with the model. If the fluctuative scaling law model is a valid hypothesis then there must be some observable feature by which those preferred parameters could be distinguished.

In the following Section, it is proposed that the characteristic parameters of the observable universe are determined by the cosmological constant (i.e. the present observed value of dark energy), resulting in certain scaling relationships between the cosmological constant and the microscopic parameters. It is useful first to establish the generalized relationships between the observable universe and the particle or particles that determine its dynamical parameters, according to the fluctuative scaling law model. Let it suffice, for the present purposes, to stipulate that there exists some characteristic mass $M_u$ and radius $R_u$ that are associated with the observable universe and that are scaled to the characteristic parameters of galaxies, solar systems and certain microscopic particles according to the model in [1,2].

The characteristic relaxation time $t_u$, associated with the observable universe, is of the order $(R_u/g_u)^{1/2}$, where $g_u \sim GM_u R_u^{-2}$ is the magnitude of the characteristic gravitational force and $G$ is the Newtonian gravitational coupling. The characteristic time $t_u$ is therefore of the order $(R_u^3/GM_u)^{1/2}$. The characteristic action $A_u$ of the observable universe is of the order $U_u t_u$, where $U_u \sim GM_u^2 R_u^{-1}$ is the magnitude of the gravitational potential energy associated with the particle horizon. The term $A_u$ is therefore of the order $(GM_u^3 R_u)^{1/2}$ [1,2]. According to the dark matter paradigm, the mass $M_u$ may be primarily composed of at least one, yet-unidentified species of particle. In that scenario, the nucleon would still represent approximately a granular component of the observable universe, but some species of dark matter particle would represent more accurately the predominant granular component whose parameters are related to the characteristic cosmic parameters. Let the particle whose mass constitutes the bulk of the characteristic cosmic mass $M_u$ be identified by the subscript $z$. The present analysis imposes no restrictions on the mass $M_z$ of that particle, and $z$ may be either baryonic or non-baryonic. The relationships in (1), (2) and (3) lead to certain scaling relationships between the parameters of the observable universe and the parameters of the particle $z$. It follows, from the basic principles of quantum mechanics, that the characteristic action of any fundamental particle must be of the order the Planck quantum constant $\hbar = h/(2\pi)$. The only physically meaningful measure of the size of a fundamental particle is then the Compton wavelength. Let the action $A_z$ of the particle $z$ be therefore of the order $h$ and let the characteristic radius $R_z$ be $h/(M_z c)$, the Compton length, where $c$ is the vacuum-speed of light. The characteristic time $t_n$, representing the relaxation time of the statistical time of a quantum fluctuation within $z$, must be of the order the time required for light to traverse the distance $R_z$, and therefore $t_z \sim h/(M_n c^2)$ [1].

If the fluctuative scaling law model is representative of nature, then the characteristic parameters of the particle $z$ and the observable universe must be related according to (1), (2) and (3), with $k=z$ and $j=u$. It follows from (1) and (2), and separately from (2) and (3), that $M_z$ must be related to $M_u$ according to

$$M_z \sim \left(\frac{h^2 c^2}{G^2 M_u}\right)^{1/3}, \qquad (4)$$

where $m_P = (hc/G)^{1/2}$ is the Planck mass. Additionally, it follows from (1) and (3) that

$$\frac{M_u}{R_u} \sim \frac{c^2}{G}. \qquad (5)$$

If the characteristic mass $M_u$ could be determined, then the mass $M_z$ of the dominant granular component could be ascertained. Furthermore, the relationship in (5) imposes a restriction on $M_u$ and $R_u$.

## 2. The fluctuative scaling law model and the cosmological constant

In a universe in which the cosmological constant $\Lambda$ is equal to zero, the only possible measure of the size of the universe is the particle horizon, since the event horizon would be infinite. The particle horizon $R_p(T)$, associated with some proper time $T$, represents the conformal distance between a given point $A$ and the most remote source of any particles that could be detected at that time by an observer located at $A$. The associated mass $M_p(T) \equiv \rho(T) R_p^3(T) 4\pi/3$, where $\rho(T)$ is the average density of matter in the observable universe, is the mass contained within the sphere whose radius is $R_p(T)$, or the "particle-sphere". The era of radiation-dominance ended when the proper time was $T_r \sim 10^{12}$s, and the universe then became matter-dominated. If $\Lambda=0$ then the universe must be matter-dominated for all times after $T_r$, and there would be consequently no other distinctive set of physical parameters other than those associated with $R_p(T_r)$. However, it would not be physically meaningful for the fluctuative scaling law model to apply to the parameters associated with the age in which $T \sim T_r$ since it is not possible to characterize the observable universe as an ensemble of stars and galaxies in such an early age. The fluctuative scaling law model would be meaningless in a universe where $\Lambda=0$ so the observed cosmological constant, different from zero, could be a fundamental probe for the model. In fact, the observed positive cosmological constant can be associated with a physically significant set of cosmological parameters whose significance is ascertained empirically (see, e.g., the results of WMAP collaboration [12]). For $\Lambda>0$, then there exists a finite event horizon $R_e(T)$ that approaches asymptotically the de Sitter horizon

$$R_\Lambda \equiv c \left(\frac{3}{\Lambda}\right)^{1/2}. \tag{6}$$

$R_e(T)$ becomes of the order $R_\Lambda$ when the proper time $T$ is of the order the fundamental cosmological time $T_\Lambda$, where

$$T_\Lambda \equiv \Lambda^{-1/2}. \tag{7}$$

The mass $M_e(T) \equiv \rho(T) R_e^3(T) 4\pi/3$, contained within the event-sphere, is never greater than the fundamental mass $M_\Lambda$ given by

$$M_\Lambda = A \frac{c^3}{G \Lambda^{1/2}}, \tag{8}$$

where $A$ is a constant of the order 20 and follows from numerical integration (for the present purposes, the coefficient $A$ and other geometrical or numerical coefficients may be ignored.). Contrary to $R_e(T)$, $M_e(T)$ approaches asymptotically $M_\Lambda$ in the limit of vanishing $T$, and decreases with increasing $T$. $M_e(T)$ remains of order near $M_\Lambda$ until $T$ becomes much greater than $T_\Lambda$, but will nearly vanish in the distant future. A positive cosmological constant does not effect any limits on the particle horizon $R_p(T)$, which will grow without bound if $\Lambda=0$ or if $\Lambda>0$. However, if $\Lambda>0$ then the mass $M_p(T)$ contained within the particle-sphere approaches asymptotically $M_\Lambda$. The term $M_p(T)$ becomes of the order $M_\Lambda$ when $T \sim T_\Lambda$.

In Fig. 1, are displayed $M_p(T)$ and $M_e(T)$ as a function of proper time $T$, and Fig. 2 presents the analogous plots for $R_p(T)$ and $R_e(T)$. During the age in which the proper time $T$ is of the order $T_\Lambda$, the horizons $R_p(T)$ and $R_e(T)$ are scaled mutually and are of the order $R_\Lambda$. On the other hand, $M_p(T)$ and $M_e(T)$ are scaled mutually and are of the order $M_\Lambda$.

It is plausible that the cosmological parameters $T_\Lambda$, $R_\Lambda$ and $M_\Lambda$ represent the characteristic parameters of the observable universe that are preferred by the fluctuative scaling law model. If $M_u \sim M_\Lambda$ and $R_u \sim R_\Lambda$ then it follows from (5) that

$$M_z \sim m_\Lambda \equiv \left(\frac{\hbar^4 \Lambda}{G^2 c^2}\right)^{1/6}. \tag{9}$$

The fluctuative scaling law model establishes therefore a scaling relationship between the mass of the predominant microscopic component of the observable universe and the cosmological constant. This relation is naturally achieved and it is based on the only hypothesis that a fundamental quantum action gives rise to a stochastic mechanism of aggregation by gravitational interaction.

## *3. Conclusions*

According to the current observations, there exists, throughout the observable universe, a vacuum-energy related to the observed positive cosmological constant. Specifically, detailed observations indicate that the cosmological constant, considered responsible for the apparent accelerated behavior of the Hubble fluid, is approximately $1.2 \times 10^{-35}\text{s}^{-2}$ [12]. The time $T_\Lambda$ is thus approximately $2.9 \times 10^{17}$s, and the current proper time $T_0$ is approximately $4.3 \times 10^{17}$s. Consequently, the current age is apparently the characteristic age associated with $\Lambda$. That remarkable fact is known as the "cosmic coincidence". Vacuum dominance began apparently early (soon after the formation of galaxies), and the universe was no longer matter-dominated. However, since $T_0 \sim T_\Lambda$, the average density of energy, associated with matter, is still of the order of the density of the vacuum-energy (this means that considering the density parameters, we have $\Omega_M \sim 0.3$ and $\Omega_\Lambda \sim 0.7$). The de Sitter horizon $R_\Lambda$ is approximately $1.5 \times 10^{26}$m and the mass $M_\Lambda$ is approximately $2.2 \times 10^{54}$kg. The event horizon $R_e \equiv R_e(T_0)$ is approximately $1.4 \times 10^{26}$m, and the particle horizon $R_p \equiv R_p(T_0)$, which is also a function of $\Lambda$, is approximately $4.4 \times 10^{26}$m. The mass $M_e \equiv M_e(T_0)$ contained within the event-sphere is approximately $3.3 \times 10^{52}$kg, and the mass $M_p \equiv M_p(T_0)$ contained within the particle-sphere is approximately $9.2 \times 10^{53}$kg. The similarities between $M_e$ and $M_p$ and between $R_e$ and $R_p$ are the result of the cosmic coincidence $T_0 \sim T_\Lambda$, which ensures that $M_p \sim M_e \sim M_\Lambda$ and $R_p \sim R_e \sim R_\Lambda$.

If $\Lambda$ is approximately $1.2 \times 10^{-35}\text{s}^{-2}$ then it follows from (9) that $M_z$ must be of order near $10^{-28}$kg. The mass of the dominant species of particle must be therefore of order near the nucleon mass! It is important to note that $M_z \sim m_\Lambda$ does not require that the particle z be identical to the nucleon or even baryonic. However, it does suggest that, since the nucleon represents approximately a granular component of the observable universe, the mass $M_n$ of the nucleon is scaled to $m_\Lambda$. That result is significant since it is consistent with a number of independent arguments. Zel'dovich first proposed that $\Lambda \propto M_n^6$ follows from basic considerations of field theory [8]. Holographic principles may also require that $M_n \sim m_\Lambda$ [9]. The scaling relationship $M_n \sim m_\Lambda$ would resolve also a number of problematic large-number coincidences among the parameters of nature [7]. Furthermore, $M_n \sim m_\Lambda$ follows from applying the Bekenstein-Hawking bound to a model for the origin of the universe in which three dimensions inflated from the collapse of seven extra dimensions [10]. In conclusion, it seems that the fluctuative scaling law model could naturally relate the fundamental granular component, the astrophysical self-gravitating systems and the cosmological constant in a consistent, at least phenomenological, scheme.

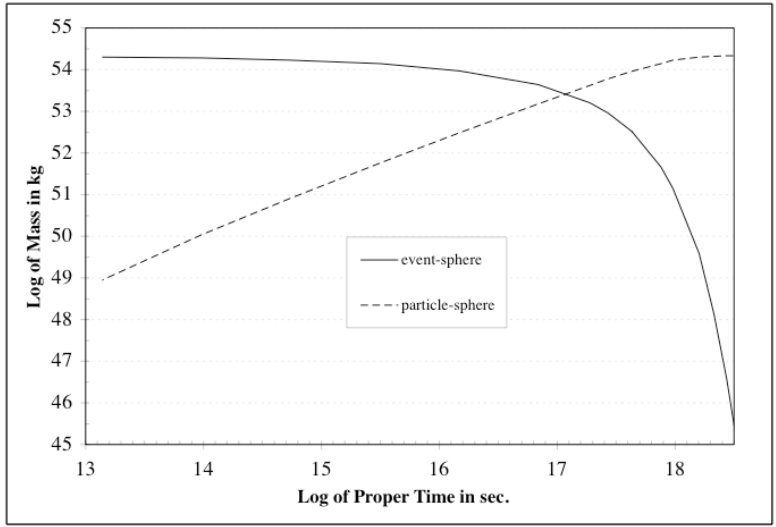

Figure 1. The log of the mass $M_e(T)$ contained within the event-sphere and the mass $M_p(T)$ contained within the particle-sphere are shown as functions of the log of proper time $T$.

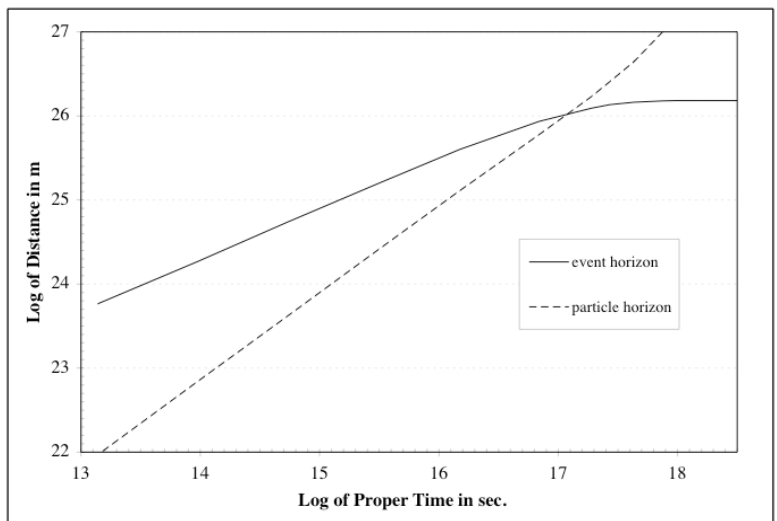

Figure 2. The log of the event-horizon $R_e(T)$ and the particle horizon $R_p(T)$ are shown as functions of the log of the proper time $T$.